\documentclass[aps,prl,twocolumn,superscriptaddress,floatfix,10pt]{revtex4-1}
\usepackage{graphicx}
\usepackage{latexsym}
\usepackage{amssymb}
\usepackage{amsmath}
\usepackage{amsfonts}
\usepackage{gensymb}
\usepackage{bm}
\usepackage{multirow}
\usepackage{color}
\usepackage{manfnt,wasysym}
\usepackage{hyperref} 

\newcommand{\nn}{\nonumber}

\newcommand{\ket}[1]{|#1 \protect\rangle}

\newcommand{\brakets}[2]{\langle #2 | #1 | #2 \rangle}

\newcommand{\RR}{\mathbb{R}}
\newcommand{\ZZ}{\mathbb{Z}}

\newcommand{\refcite}[1]{Ref.\,\cite{#1}}
\newcommand{\eqnref}[1]{Eq.\,(\ref{#1})}
\newcommand{\figref}[1]{Fig.\,\ref{#1}}

\newcommand{\secref}[1]{Sec.\,\ref{#1}}
\newcommand{\appref}[1]{Appendix~\ref{#1}}
\newcommand{\mysec}[1]{\section{#1}}
\newcommand{\citeapp}[1]{(\appref{#1})}

\newcommand{\cube}{\text{\mancube}}

\definecolor{myGreen}{RGB}{0,204,0}
\newcommand{\cP}{\text{\textcolor{myGreen}{\bf{green}}}}
\newcommand{\cK}{\text{\bf{black}}}

\newif\ifJournal
\Journaltrue 

\ifJournal
\newcommand{\foot}[1]{\footnote{#1}}
\newcommand{\citefoot}[1]{\cite{#1}}
\newcommand{\sfigref}[2]{Fig.\,\hyperref[#1]{\ref{#1}(#2)}}
\else
\newcommand{\foot}[1]{[\footnote{#1}]}
\newcommand{\citefoot}[1]{[\cite{#1}]}
\newcommand{\sfigref}[2]{Fig.\,\hyperref[#1]{\ref{#1}#2}}
\fi

\newcommand{\daggerNote}{(Regarding daggers in figure: \citefoot{foot:daggers}.)}

\hypersetup{
  colorlinks = true,
  urlcolor = blue,
  pdfauthor = {Kevin Slagle, Yong Baek Kim},
  pdftitle = {Slagle and Kim - 2018 - X-cube model on generic lattices: Fracton phases and geometric order}
}

\begin{document}

\title{X-cube model on generic lattices: Fracton phases and geometric order}

\author{Kevin Slagle}
\affiliation{Department of Physics, University of Toronto, Toronto, Ontario M5S 1A7, Canada}

\author{Yong Baek Kim}
\affiliation{Department of Physics, University of Toronto, Toronto, Ontario M5S 1A7, Canada}
\affiliation{Canadian Institute for Advanced Research, Toronto, Ontario, M5G 1Z8, Canada}

\begin{abstract}
Fracton order is a new kind of quantum order characterized by topological excitations that exhibit remarkable mobility restrictions
  and a robust ground state degeneracy (GSD) which can increase exponentially with system size.
In this paper, we present a generic lattice construction (in three dimensions) for a generalized X-cube model of fracton order,
  where the mobility restrictions of the subdimensional particles inherit the geometry of the lattice.
This helps explain a previous result that lattice curvature can produce a robust GSD, even on a manifold with trivial topology.
We provide explicit examples to show that the (zero temperature) phase of matter is sensitive to the lattice geometry.
In one example, the lattice geometry confines the dimension-1 particles to small loops,
  which allows the fractons to be fully mobile charges,
  and the resulting phase is equivalent to (3+1)-dimensional toric code.
However, the phase is sensitive to more than just lattice curvature;
  different lattices without curvature (e.g. cubic or stacked kagome lattices) also result in different phases of matter,
  which are separated by phase transitions.
Unintuitively however, according to a previous definition of phase in \refcite{XiePhases},
  even just a rotated or rescaled cubic lattice results in different phases of matter,
  which motivates us to propose a new and coarser definition of phase for gapped ground states and fracton order.
The new equivalence relation between ground states is given by the composition of a local unitary transformation \emph{and} a quasi-isometry (which can rotate and rescale the lattice);
  equivalently, ground states are in the same phase if they can be adiabatically connected by varying both the Hamiltonian \emph{and} the positions of the degrees of freedom (via a quasi-isometry).
In light of the importance of geometry,
  we further propose that fracton orders should be regarded as a \emph{geometric order}.
\end{abstract}


\maketitle

Topologically ordered quantum phases of matter are often characterized by their topological excitations
  and topological ground state degeneracy (GSD).
These (liquid \cite{ZengLiquids}) topological phases \cite{KitaevHoneycomb,Wen2D,Lan2017} (e.g. $Z_2$ gauge theory described by toric code \cite{KitaevToric} or BF theory \cite{BFTheory,PutrovBraiding}) are topologically invariant.
That is, the Lagrangian has no dependence on the spacetime metric;
  the GSD only depends on the topology of the spatial manifold;
  and the braiding statistics of the topological excitations only depends on the topology of the braiding paths.

\begin{figure}
\begin{minipage}{.3\columnwidth}
\includegraphics[width=\columnwidth]{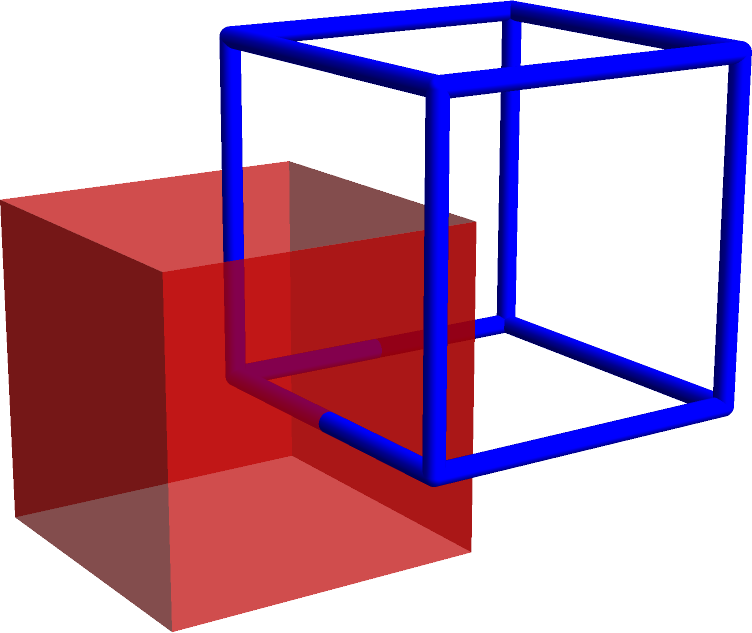}
\end{minipage}
\hspace{.02\columnwidth}
\begin{minipage}{.3\columnwidth}
\includegraphics[width=\columnwidth]{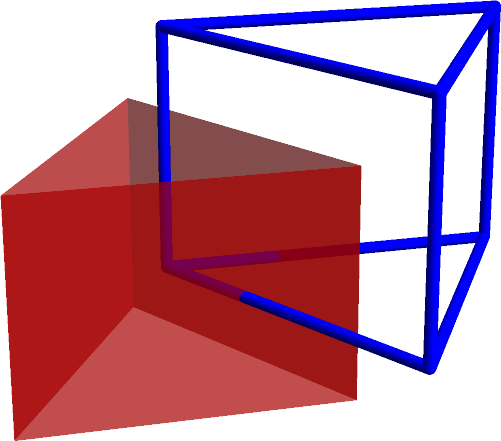}
\end{minipage}
\hspace{.02\columnwidth}
\begin{minipage}{.3\columnwidth}
\includegraphics[width=\columnwidth]{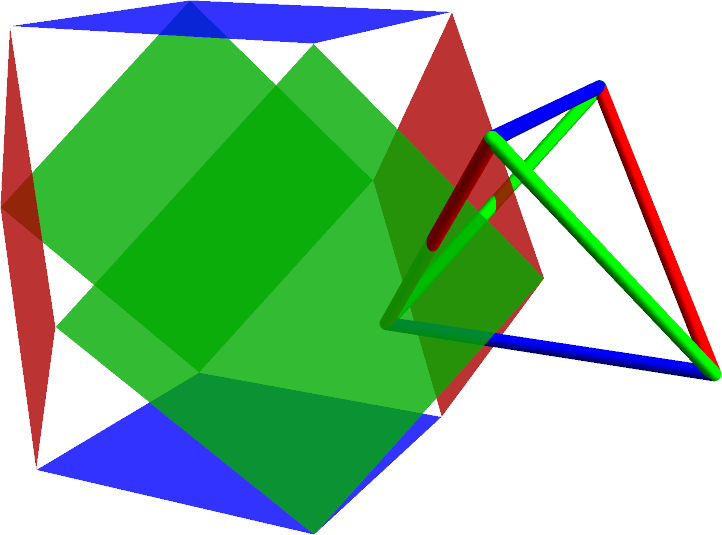}
\end{minipage} \\
\begin{minipage}{.3\columnwidth}
\textbf{(a)}
\end{minipage}
\hspace{.02\columnwidth}
\begin{minipage}{.3\columnwidth}
\textbf{(b)}
\end{minipage}
\hspace{.02\columnwidth}
\begin{minipage}{.3\columnwidth}
\textbf{(c)}
\end{minipage}
\caption{
Geometric braiding operators for the
  \textbf{(a)} X-cube model on a cubic lattice \cite{VijayXCube,MaLayers,fractonQFT},
  \textbf{(b)} X-cube model on a stacked kagome lattice,
  and \textbf{(c)} Chamon model \cite{ChamonModel,Bravyi2011} of fracton order.
The surfaces are membrane operators that braid fractons around the edge of the membrane by exchanging other excitations in the interior of the membrane \cite{fractonQFT}.
The lines are string operators that braid dimension-1 particles.
The operators are products of $X$, $Y$, and $Z$ Pauli operators,
  which are colored red, green, and blue.
These operators can be used to detect subdimensional particles.
This figure exemplifies that in fracton orders,
  the braiding paths of the subdimensional excitations take rigid geometric shapes that depend on the model and lattice.
}\label{fig:braid}
\end{figure}

This topological invariance is absent in the recently discovered, exactly solvable, gapped three-dimensional lattice models
  \cite{VijayFracton,HaahCode,Yoshida2013,ChamonModel,Bravyi2011,VijayXCube,VijayNonabelian,MaLayers,HsiehPartons,Hsieh2017,Petrova_Regnault_2017,Brown2016}
  that exhibit so-called fracton topological order \cite{VijayXCube}
  \foot{However, since fracton orders are not topologically invariant, including the word ``topological'' in ``fracton topological order'' is misleading.
        We propose that fracton geometric order would be a more accurate name.}.
Fracton order can be characterized by its topological excitations
  which are subdimensional \cite{PretkoU1},
  which means that they are immobile or are restricted to only move along lines or surfaces 
    without creating or destroying other topological excitations.
In this context, a topological excitation is an excitation that can not be annihilated by local operators,
  but instead requires contact with a corresponding antiparticle in order to be annihilated.
  \foot{Other examples of topological excitations include ordinary electrons and the charge and flux excitations of toric code \cite{KitaevToric}.}
The immobile excitations are called fractons,
  while the particles that are bound to lines and surfaces are called dimension-1 particles (or lineons \cite{Devakul_Parameswaran_Sondhi_2017})
  and dimension-2 particles, respectively.
Fracton order has also been characterized by its GSD which increases exponentially with system size on a torus \cite{VijayFracton,HaahCode,XcubeRG},
  geometric braiding processes (\figref{fig:braid}),
  geometry-dependent entanglement \cite{Shi2017,Ma_2017,Schmitz_2017},
  glassy dynamics \cite{PremHaahNandkishore,ChamonModel,PretkoTemperature},
  duality to lattice defects \cite{PretkoDuality},
  duality to fractal and subdimensional symmetry breaking \cite{Yoshida2013,VijayXCube,WilliamsonUngauging,Slagle2spin,NUSSINOV2009977,Nussinov06102009},
  bifurcation in entanglement renormalization \cite{Haah2014,XcubeRG},
  and connections to emergent gravity \cite{PretkoGravity}.

\begin{figure}
\includegraphics[width=.43\columnwidth]{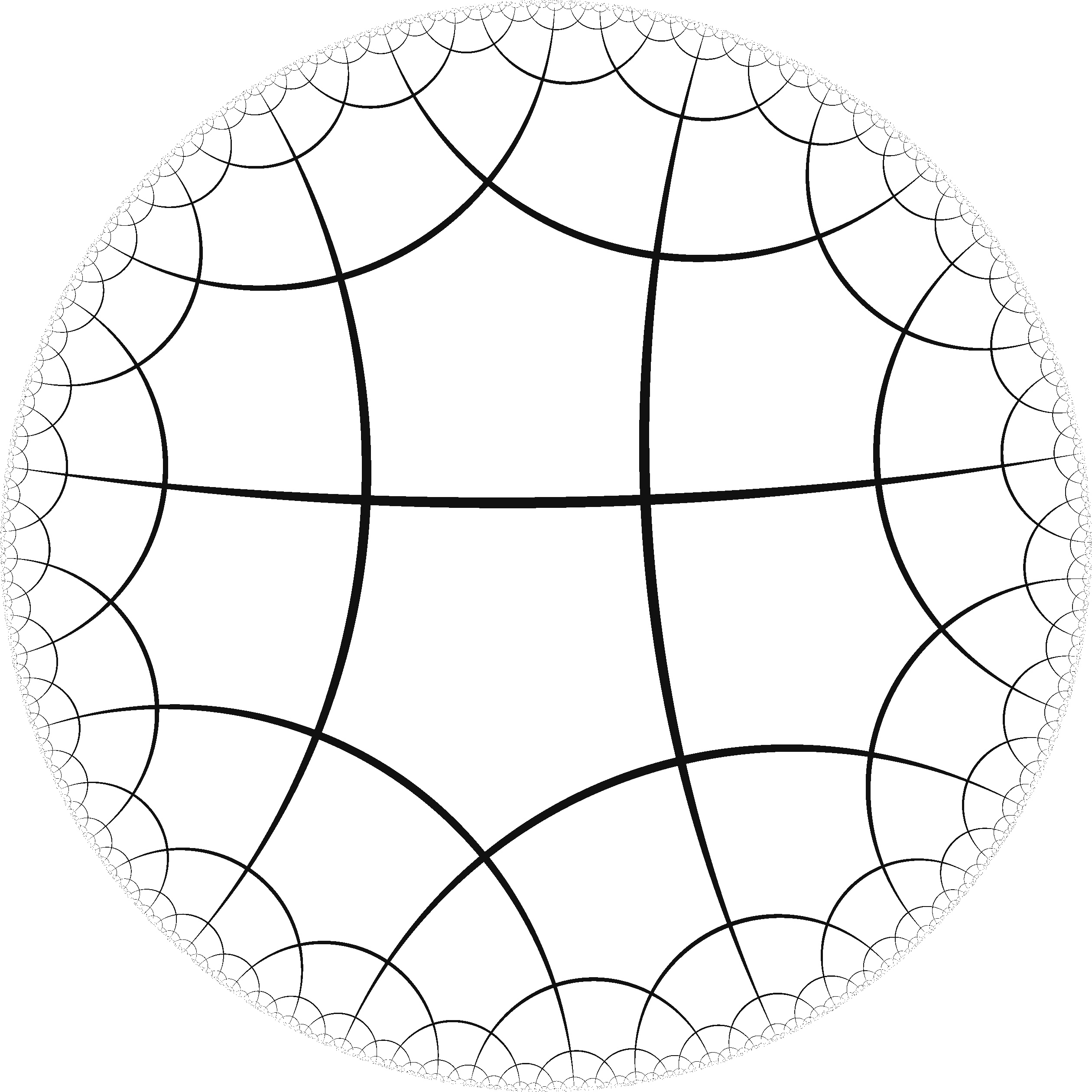}
\caption{
Order-4 pentagonal tiling of the hyperbolic plane.
A stack of this lattice can be created from the intersecting surface construction.
The dimension-1 particles move along the lines of the above lattice, which are geodesics of the hyperbolic plane.
As a result, the braiding operators in \figref{fig:braid} will curve with the lattice geometry.
The geometry of a hyperbolic 3-space can be constructed from an order-4 dodecahedral honeycomb \cite{dodecahedralHoneycomb}.
}\label{fig:hyperbolic}
\end{figure}

The aim of our work is to demonstrate the significance of geometry in the study of fracton order
  by considering the effects of changing the geometry of the lattice.
By lattice geometry, we mean the following: Is the lattice a cubic lattice or stack of kagome lattices?
  Or does the lattice have curvature (as in e.g. \figref{fig:hyperbolic})?
Although liquid topological order (e.g. $Z_2$ gauge theory) is completely blind to lattice geometry,
  we will show that lattice geometry plays a fundamental role in the physics of fracton order.
To do this, in \secref{sec:isurface} we formulate a lattice construction (\figref{fig:cubicKagome})
  of generic lattices on which we can define a generalized X-cube fracton model,
  which was previously only defined on a cubic lattice \cite{VijayXCube}.

\begin{figure}
\includegraphics[width=.4\columnwidth]{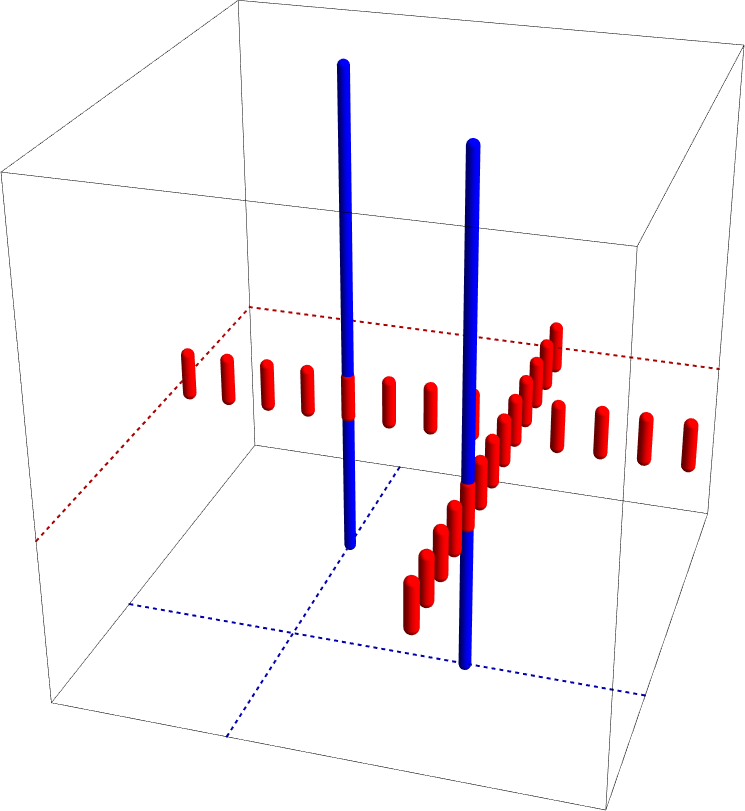}
\hspace{.5cm}
\includegraphics[width=.45\columnwidth]{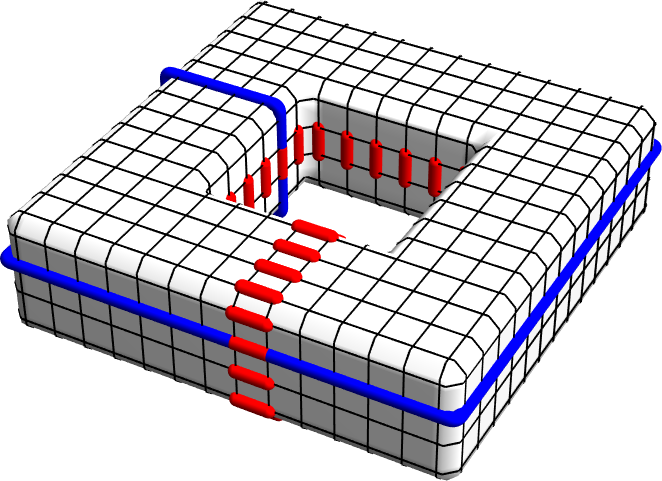}
\begin{minipage}{.4\columnwidth}
\textbf{(a)}
\end{minipage}
\hspace{.5cm}
\begin{minipage}{.45\columnwidth}
\textbf{(b)}
\end{minipage}
\caption{
\textbf{(a)} Wilson (blue) and 't Hooft loops (red) of the X-cube model on a cubic lattice,
  which are the logical operators that act on the degenerate ground state manifold \cite{fractonQFT,He_2017}.
These loops are products of $Z$ (blue) and $X$ (red) operators.
Physically, a Wilson loop moves a dimension-1 particle,
  while a 't Hooft loop moves a dimension-2 pair of fractons.
\textbf{(b)} When a large torus-shaped i-surface is added to a cubic lattice (hidden for clarity),
  new links (black) with new qubits are added to the model.
New Wilson and 't Hooft loops result on the torus-shape i-surface,
  which results in a GSD that can be attributed to lattice curvature. \cite{fractonQFT}
}\label{fig:loopsAndDegen}
\end{figure}

In \secref{sec:geometry} we will see that the mobility restrictions of the subdimensional particles inherit the geometry of the lattice.
This helps explain a previous discovery that
  lattice curvature can result in a robust GSD on a manifold with trivial topology (\sfigref{fig:loopsAndDegen}{b}) \cite{fractonQFT}.

In \secref{sec:mobilizing}, we show that the lattice geometry can also affect the phase of matter (at zero temperature).
As examples, we consider two lattices (\figref{fig:spheres} and \ref{fig:circles}) where the geometry grants fractons
  either full or subdimensional mobility which results in a phase equivalent to 3+1D $Z_2$ gauge theory or a stack of 2+1D toric codes.

In \secref{sec:phases}, we show that the phase is sensitive to more than just lattice curvature.
For example, the X-cube model on different lattices without curvature (e.g. cubic or stacked kagome)
  can result in a different GSD and phase of matter.

In \refcite{XiePhases} (and \cite{Hastings_2010}), two gapped quantum ground states were defined to be in the same phase if they could be connected by an adiabatic evolution of the Hamiltonian.
Equivalently, the phase was also classified by an equivalence relation where two states are in the same phase if they could be connected by a (generalized) local unitary (gLU) transformation.
(Hamiltonians on different lattices can be compared by adding trivial gapped degrees of freedom so that both Hamiltonians share the same Hilbert space;
  analogously, a ``generalized'' local unitary is allowed to add and remove degrees of freedom that are in a direct product state.)

Remarkably, we show that using this definition of phase \cite{XiePhases},
  a rotated or rescaled lattice can also correspond to a different phase (\figref{fig:phases});
  e.g. the X-cube model on cubic lattices with different orientations corresponds to different phases of matter.
This surprising (and unintuitive) result can be understood from the fact that different lattices have different
  lines and surfaces that the subdimensional particles are bound to (\figref{fig:dim2ops}).

This motivates us to propose a coarser definition of phase in \secref{sec:isometry} which (more intuitively) equates the X-cube model on rotated and rescaled lattices.
Under the new definition, two gapped quantum ground states are in the same phase if they can be connected by an adiabatic evolution of both the Hamiltonian \emph{and} the positions of the degrees of freedom.
Equivalently, the new definition of phase is also given by an equivalence relation where two states are in the same phase if they can be connected by the composition of a generalized local unitary (gLU) transformation \emph{and} a quasi-isometry (which can e.g. rotate and rescale the lattice).
A quasi-isometry is a spatial transformation that preserves long-distance structure, such as locality,
  but is not required to preserve short-distance structure.
When only liquid phases such as liquid \cite{ZengLiquids} topological order are considered,
  our new definition of phase reduces to the previous definition proposed in \refcite{XiePhases}.

\begin{figure}
\begin{minipage}{.24\columnwidth}
\includegraphics[width=\columnwidth]{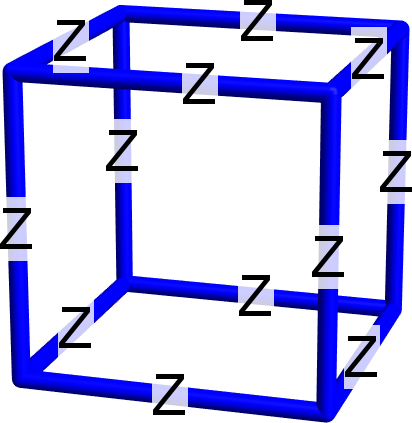} \\
\textbf{(a)}
\end{minipage}
\begin{minipage}{.74\columnwidth}
\includegraphics[width=.32\columnwidth]{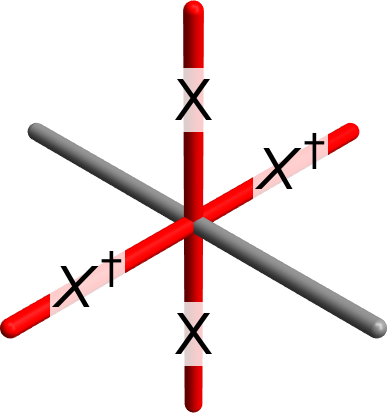}
\includegraphics[width=.32\columnwidth]{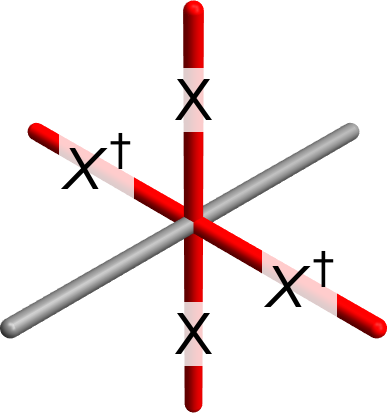}
\includegraphics[width=.32\columnwidth]{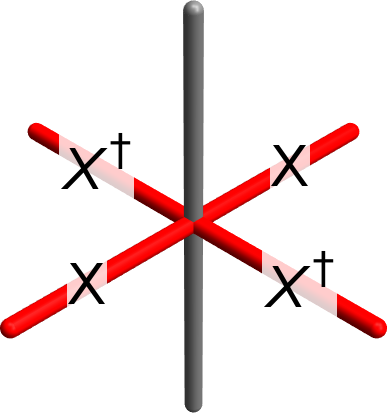} \\
\textbf{(b)}
\end{minipage}
\caption{
\textbf{(a)} For every cube, the X-cube model on a cubic lattice (\eqnref{eq:Xcube H}) has a 3-cell operator which is a product of 12 Pauli $Z$ operators along the edges of the cube: $\prod_{\ell \in \cube} Z_\ell$.
\textbf{(b)} At each vertex, there are three cross operators:
  one for each of the three planes that intersect the vertex.
The operators are a product of four $X$ operators on the four links within the plane that neighbor the vertex: $\prod_{\ell \in +} X_\ell$.
\daggerNote
}\label{fig:Xcube}
\end{figure}

\begin{figure}
\begin{minipage}{.34\columnwidth}
\includegraphics[width=\columnwidth]{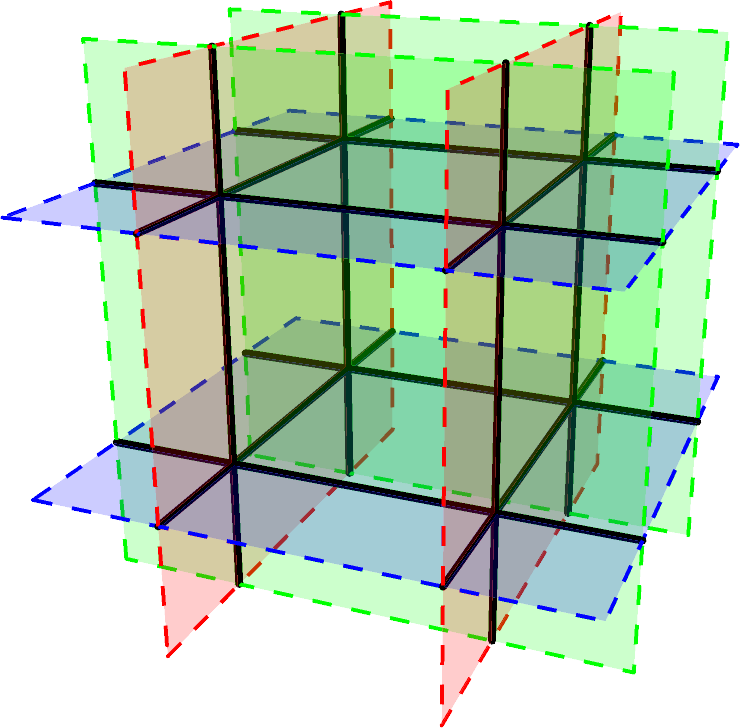}
\end{minipage}
\begin{minipage}{.64\columnwidth}
\includegraphics[width=\columnwidth]{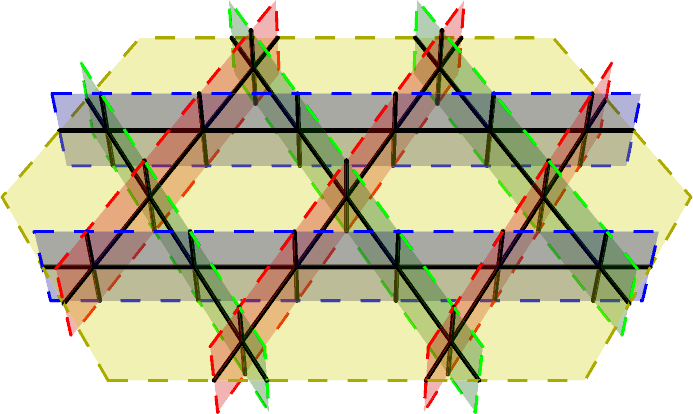}
\end{minipage} \\
\begin{minipage}{.34\columnwidth}
\textbf{(a)}
\end{minipage}
\begin{minipage}{.64\columnwidth}
\textbf{(b)}
\end{minipage}
\caption{
\textbf{(a)} A cubic lattice and \textbf{(b)} a stack of kagome lattices constructed from intersecting surfaces (colored planes),
  which we refer to as i-surfaces. 
Pauli operators live on the links (black) of the lattice,
  which are placed where two i-surfaces intersect.
The Hamiltonian (\eqnref{eq:Xcube H'}) consists of three cross operators at each vertex (\sfigref{fig:Xcube}{b}),
  and a 3-cell operator at each 3-cell (3D volume enclosed by i-surfaces) which is a product of $Z$ operators on the edges of the 3-cell.
The stack of kagome lattices has two kinds of 3-cells: a triangular prism (\sfigref{fig:3cell}{a}) and a hexagonal prism.
}\label{fig:cubicKagome}
\end{figure}

\mysec{X-Cube Model Review} \label{sec:review}
The X-cube model was originally defined on a cubic lattice with $Z_2$ Pauli operators on the links \cite{VijayXCube}:
\begin{equation}
  H_\text{X-cube} = - \sum_\cube \prod_{\ell \in \cube} Z_\ell - \sum_+ \prod_{\ell \in +} X_\ell \label{eq:Xcube H}
\end{equation}
The first term sums over all cubes in the lattice and is a product of 12 Pauli $Z$ operators over the 12 edges of the cube (\sfigref{fig:Xcube}{a}).
Excitations of this term are immobile fractons which are created at the corners of rectangular membrane operators \cite{VijayXCube}.
However, a pair of neighboring fractons is a dimension-2 particle which can move along a plane
  (via the same ``membrane'' operator but of unit width).
This cube operator counts the number of fractons within the cube
  by braiding dimension-1 particles (excitations of the second term) around the edges of the cube.

The second term in the Hamiltonian sums over all quadruples of links which make the shape of a cross and is a product of four Pauli $X$ operators over these four links (\sfigref{fig:Xcube}{b}).
Excitations of this term are dimension-1 particles which can only move along the x, y, or z axes \cite{VijayXCube}.
The collection of an x-axis, y-axis, and z-axis dimension-1 particle can fuse into the vacuum.
A neighboring pair of dimension-1 particles moving in the same direction is a dimension-2 particle,
  which can move along the plane orthogonal to their displacement.
The cross operator in the XY plane counts the number of x-axis and y-axis particles at the vertex (modulo 2)
  by braiding a pair of fractons around a loop in the XY plane.

\mysec{Intersecting Surfaces Lattice Construction} \label{sec:isurface}
Unlike liquid topological order, the X-cube model can not be naturally defined on an arbitrary lattice.
The links neighboring each vertex must come in pairs to uniquely specify how a dimension-1 particle should pass through a link.
We may also want to preserve the fusion rule that the collection of three orthogonal dimension-1 particles can fuse into the vacuum.
Thus, we will restrict the vertices to have exactly six neighboring links so that there are exactly three kinds of dimension-1 particles at each vertex.
Therefore, each vertex must locally look like the vertex of a cubic lattice.
Additionally, there must be a notion of surfaces for the dimension-2 particles to be bound to.

In order to facilitate these conditions, we will construct our lattice from a collection of intersecting surfaces,
  which we will refer to as i-surfaces. 
The motion of a dimension-2 fracton pair will be restricted to these i-surfaces.
The dimension-1 particles will traverse the lines formed by the intersection between two i-surfaces.
The lattice has a vertex wherever three i-surfaces intersect.
Links between vertices are places where two i-surfaces intersect.
As desired, not all lattices are compatible with this construction;
  e.g. a stack of honeycomb lattices cannot be constructed,
  which is sensible since it is not clear how a dimension-1 particle should pass through a 5-valence vertex of a stacked honeycomb lattice.
For simplicity, we will require that the i-surfaces are not fine tuned;
  i.e. perturbing the i-surfaces should not change the lattice.
This implies that no more than three i-surfaces can intersect at a single point
  and no more than two i-surfaces can intersect along a line (which e.g. rules out a stack of triangular lattices).
As examples, in \figref{fig:cubicKagome} we show how this construction can form a cubic lattice or stack of kagome lattices.

\begin{figure}
\begin{minipage}{.25\columnwidth}
\includegraphics[width=\columnwidth]{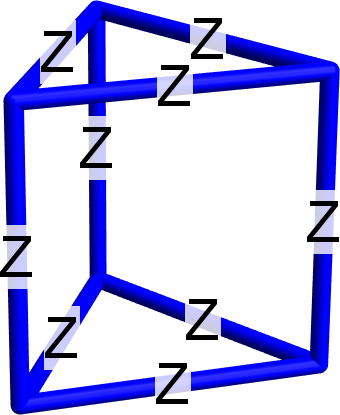}
\end{minipage}
\begin{minipage}{.35\columnwidth}
\includegraphics[width=\columnwidth]{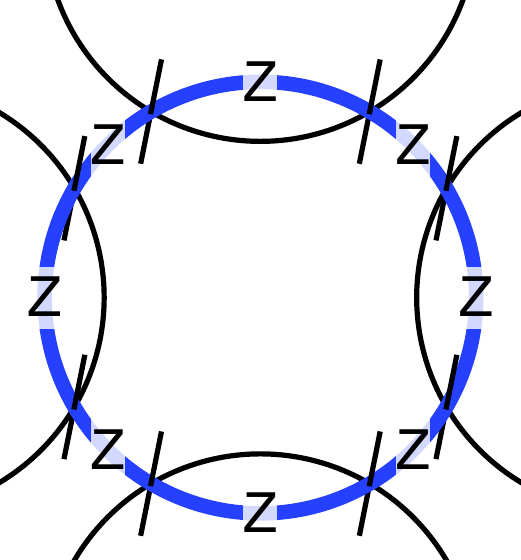}
\end{minipage}
\begin{minipage}{.35\columnwidth}
\includegraphics[width=\columnwidth]{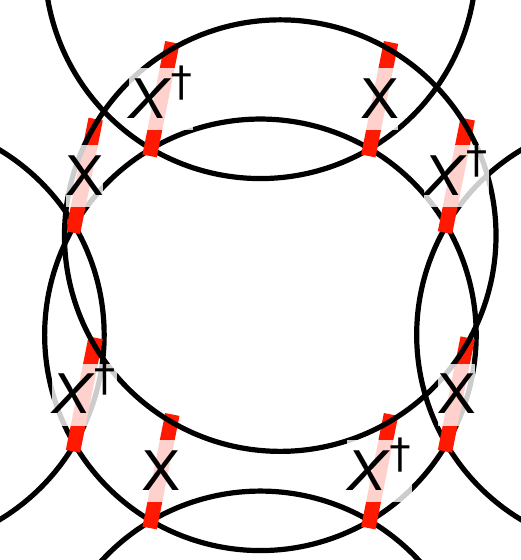}
\end{minipage}
\begin{minipage}{.25\columnwidth}
\textbf{(a)}
\end{minipage}
\begin{minipage}{.35\columnwidth}
\textbf{(b)}
\end{minipage}
\begin{minipage}{.35\columnwidth}
\textbf{(c)}
\end{minipage}
\caption{
New kinds of X-cube Hamiltonian terms in \eqnref{eq:Xcube H'}.
\textbf{(a)}
An example of a 3-cell operator on a triangular prism, which is a 3-cell in a stack of kagome lattices (\sfigref{fig:cubicKagome}{b}).
The 3-cell operator ($\prod_{\ell \in \cube} Z_\ell$) on a 3-cell ($\cube$) is a product of $Z$ operators on the links around the edges of the 3-cell.
\textbf{(b)} A product of $Z$ operators on the links around a loop: $\prod_{\ell \in \ocircle} Z_\ell$.
\textbf{(c)} A product of $X$ operators on links connecting two parallel loops: $\prod_{\ell \perp \circledcirc} X_\ell$.
\daggerNote
}\label{fig:3cell}
\end{figure}

The Hamiltonian is:
\begin{align}
\begin{split}
  H_\text{X-cube} &=     - \sum_\cube \prod_{\ell \in \cube} Z_\ell - \sum_+ \prod_{\ell \in +} X_\ell \label{eq:Xcube H'}\\
                  &\quad - \sum_\ocircle \prod_{\ell \in \ocircle} Z_\ell - \sum_\circledcirc \prod_{\ell \perp \circledcirc} X_\ell
\end{split}
\end{align}
The first two terms generalize \eqnref{eq:Xcube H}.
Instead of summing over cube operators in the first term,
  we instead sum over all 3-cells (3D volumes enclosed by i-surfaces) at which the 3-cell operator is a product of $Z$ operators on the edges of the 3-cell (\sfigref{fig:3cell}{a}).
The second term again consists of three cross operators at each vertex,
  one for each of the three i-surfaces intersecting the vertex (\sfigref{fig:Xcube}{b}).
The third term sums over all finite-sized loops (that do not increase in size as the system size increases)
  and is a product of $Z$ operators around the loop (\sfigref{fig:3cell}{b}).
The fourth term sums over all finite-sized parallel loops
  and is a product of $X$ operators on the links connecting the parallel loops (\sfigref{fig:3cell}{c}).
The last two terms are new, and only appear when there are finite-sized intersections between i-surfaces.
Without the last two terms, the model can be fine tuned (e.g. in \figref{fig:circles}).
The third term can be thought of as condensing a dimension-1 particle around the loop,
  while the fourth term condenses a (dimension-2) pair of fractons around the loop.
These terms tend to suppress the existence of immobile fracton excitations.
  \foot{
    It is possible to have 3-cells and loops that get arbitrarily large.
    In order to keep the Hamiltonian local (as defined in \eqnref{eq:local}),
      the coefficients of the 3-cell and loop operators in \eqnref{eq:Xcube H'} should decay exponentially with the size of the operator.}

\mysec{Significance of Geometry} \label{sec:geometry}
Now that we can define the X-cube model on different lattices, we can ask the following:
  How does the geometry of the lattice affect the long distance physics?

For the case of liquid topological orders, such as $Z_2$ gauge theory which is described by toric code or BF theory,
  the geometry of the lattice or curvature of the spatial manifold has little effect on the long distance physics.
That is, it does not matter if toric code is defined on a square lattice or triangular lattice;
  the charge and flux excitations can always move in any direction and the GSD only depends on the topology of the spatial manifold.

In contrast to liquid topological order, the X-cube model is very sensitive to lattice geometry.
For example, on the cubic lattice there are three kinds of dimension-1 particles,
  which are constrained to only move along the x, y, and z--axis.
However, when the X-cube model is defined on a stack of kagome lattices (\sfigref{fig:cubicKagome}{b}),
  there are \emph{four} kinds of dimension-1 particles corresponding to the four different directions that the links of the lattice are aligned.
These four kinds of dimension-1 particles are physically distinct;
  they belong to different superselection sectors and can be distinguished by braiding pairs of fractons from a distance
  (similar to how braiding can be done on a cubic lattice \sfigref{fig:braid}{a}).

As a more exotic lattice example, we can consider curved i-surfaces that produce curved lattices.
For example, a collection of curved surfaces can produce lattices with hyperbolic geometry (\figref{fig:hyperbolic}).
On this lattice, the dimension-1 particles move along curved lines, which are geodesics of the hyperbolic plane.
Thus, the mobility restrictions of the subdimensional particles inherit the geometry of the lattice.
Consequently, the rigid braiding operators (\sfigref{fig:braid}{a-b}) also inherit the lattice geometry.

The geometry-dependent mobility restrictions of the subdimensional particles also affects the ground state degeneracy (GSD).
Similar to toric code, the GSD of the X-cube model can be understood as resulting from
  non-local logical operators that act on the degenerate ground state Hilbert space.
These non-local operators are anticommuting Wilson and 't Hooft loops
  which correspond to moving a dimension-1 particle or a dimension-2 fracton pair, respectively, around a closed loop (\sfigref{fig:loopsAndDegen}{a}) \cite{fractonQFT,He_2017}.
  \foot{The non-local logical operators for toric code are also Wilson and 't Hooft loops which correspond to moving charges and fluxes, respectively, around non-contractible closed loops.}
Thus, the Wilson and 't Hooft loops share the geometric mobility constraints of the subdimensional particles.
Since the 't Hooft loops are bound to a 2-dimensional i-surface, they are very similar to 't Hooft loops in 2D toric code.
As a result, the GSD typically scales as
\begin{equation}
  \log_2\text{GSD} \sim \sum_s 2 g_s \label{eq:degen}
\end{equation}
where $\sum_s$ sums over all i-surfaces which each contribute a factor of $2^{2 g_s}$ to the GSD,
  where $g_s$ is the genus of the i-surface, $s$
  \foot{If the i-surface $s$ in \eqnref{eq:degen} is not connected and orientable, then $2 g_s$ should be replaced by the first Betti number with $Z_2$ coefficients.}.

The geometry dependence of the Wilson and 't Hooft loops allows us to better understand a previous result that
  lattice curvature can lead to a robust GSD on a manifold with trivial topology \cite{fractonQFT}.
The lattice considered is formed by constructing the cubic lattice from orthogonal i-surfaces (\sfigref{fig:cubicKagome}{a})
  and then adding an additional large i-surface with the topology of a torus (\sfigref{fig:loopsAndDegen}{b}).
The new torus-shaped i-surface has genus $g=1$
  and results in closed Wilson and 't Hooft loop operators around the torus,
  which contributes a factor of four to the GSD (in accordance with \eqnref{eq:degen}).
The GSD is robust in the limit of a large torus-shaped i-surface.
Note that the topology of the spatial manifold wasn't changed;
  instead, it was argued in \refcite{fractonQFT} that the resulting lattice should be interpreted as having spatial curvature
  around the torus-shaped i-surface.

\mysec{Mobilizing Fractons} \label{sec:mobilizing}

We will now consider two lattices with a very large amount of positive curvature.

\subsection{Cylindrical i-surfaces}

\begin{figure}
\includegraphics[width=.75\columnwidth]{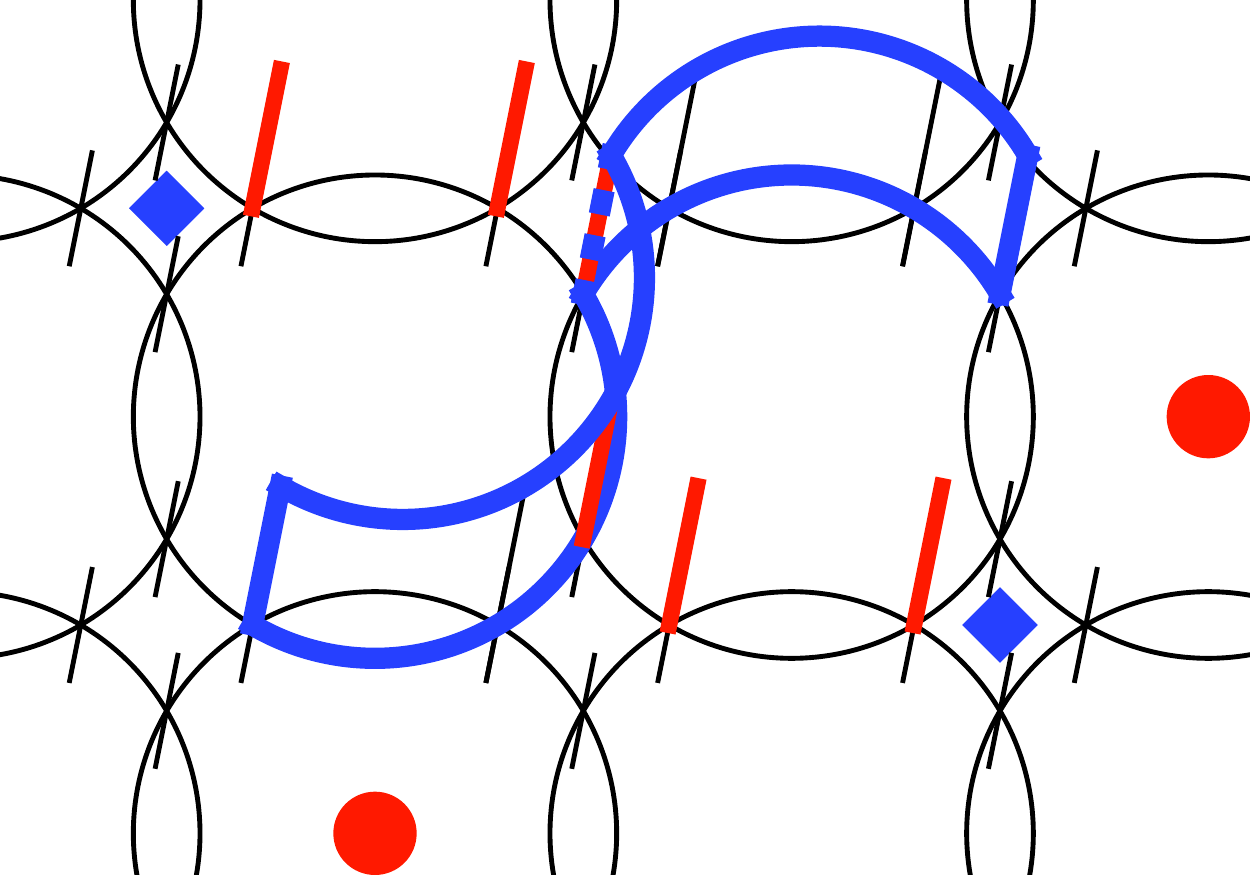}
\caption{
The X-cube model on a stack of the above lattice results in a phase equivalent to a stack of 2+1D toric codes.
Toric code charges (blue diamonds) and fluxes (red disks) are created at the ends of string operators
  which are products of $X$ (red lines) and $Z$ (blue lines) operators, respectively.
A charge is an excitation of a cube-shaped 3-cell (e.g. at a blue diamond).
A flux is an excitation of cross (\sfigref{fig:Xcube}{b}) and loop (\sfigref{fig:3cell}{c}) operators.
There are no fractons;
  the lattice curvature granted fractons mobility and turned them into toric code charges.
}\label{fig:circles}
\end{figure}

The first lattice is a stack of the lattice shown in \figref{fig:circles},
  which is constructed from cylinder-shaped i-surfaces.
Note that on this lattice, we must also include the new loop terms in $H_\text{X-cube}$ (\eqnref{eq:Xcube H'}).
These terms do not commute with Wilson and 't Hooft string operators (\sfigref{fig:loopsAndDegen}{a}) that are orthogonal to the plane,
  which implies that the topological excitations can not move orthogonal to the plane.

More physically, the third term in \eqnref{eq:Xcube H'} condenses the dimension-1 around loops shown in \sfigref{fig:3cell}{b};
  this prevents pairs of fractons from moving out of the plane.
The fourth term in \eqnref{eq:Xcube H'} condenses pairs of fractons around loops shown in \sfigref{fig:3cell}{c};
  this prevents dimension-1 particles from moving out of the plane.
These confinement processes also result in the mobilization of the fractons,
  which typically results in our lattice construction when three i-surfaces intersect at multiple points.

The resulting phase is equivalent to a stack of 2+1D toric codes.
String operators that move the toric code charges and fluxes are shown in \figref{fig:circles}.

It is important to note that the confinement processes did not result from optionally adding the third and fourth terms to $H_\text{X-cube}$ (\eqnref{eq:Xcube H'}) and increasing their strength.
Rather, the confinement processes are an instability of the first two terms of $H_\text{X-cube}$ (\eqnref{eq:Xcube H'}) that results from the geometry of the lattice,
  and in particular, the small diameter of the i-surfaces.
That is, if only the first two terms of $H_\text{X-cube}$ are considered,
  then generic perturbations will result (e.g. after applying degenerate perturbation theory)
  in an effective Hamiltonian that includes all four of the terms in $H_\text{X-cube}$.

\subsection{Spherical i-surfaces}

\begin{figure}
\begin{minipage}{.4\columnwidth}
\includegraphics[width=\columnwidth]{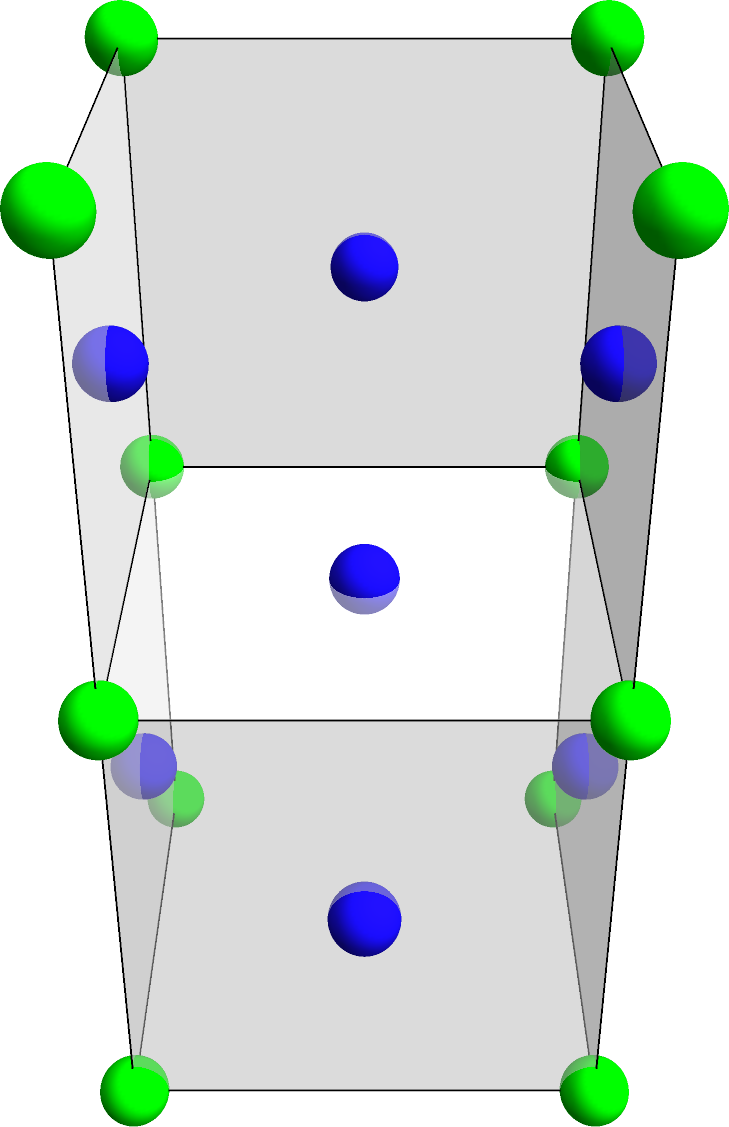}
\end{minipage}
\begin{minipage}{.54\columnwidth}
\includegraphics[width=\columnwidth]{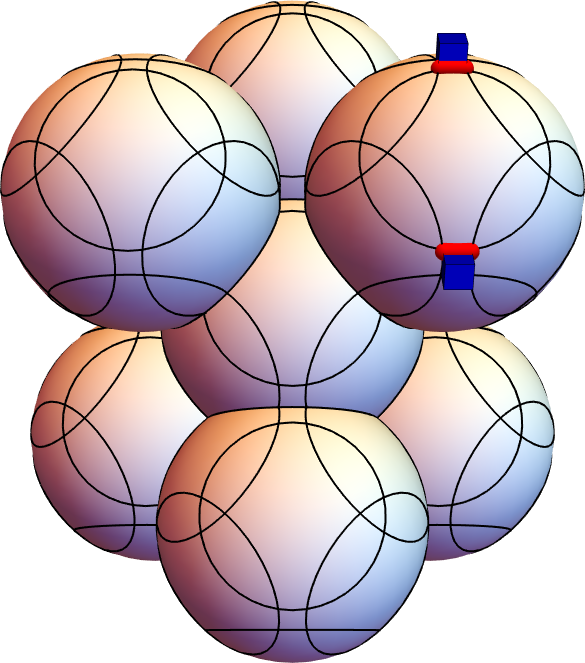}
\end{minipage} \\
\begin{minipage}{.4\columnwidth}
\textbf{(a)}
\end{minipage}
\begin{minipage}{.54\columnwidth}
\textbf{(b)}
\end{minipage} \\
\includegraphics[width=.22\columnwidth]{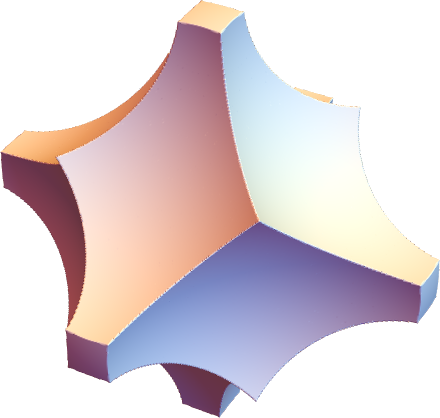}
\includegraphics[width=.21\columnwidth]{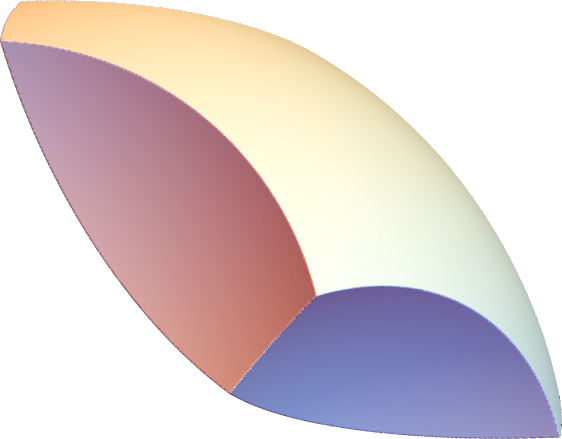}
\includegraphics[width=.14\columnwidth]{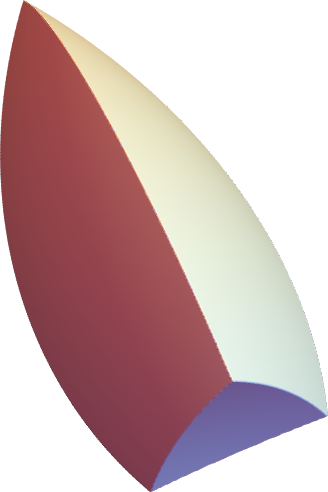}
\includegraphics[width=.19\columnwidth]{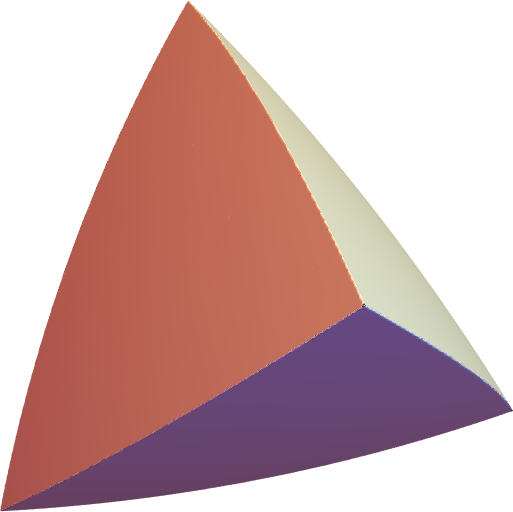}
\includegraphics[width=.19\columnwidth]{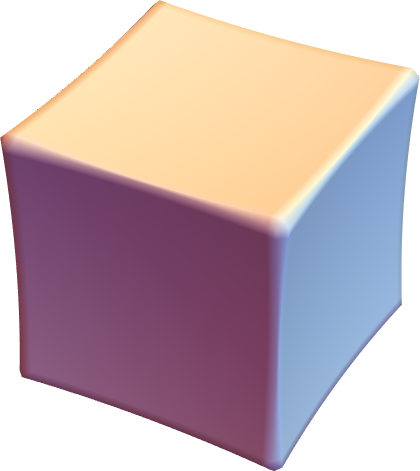} \\
\begin{minipage}{.22\columnwidth}\textbf{(c)}\end{minipage}
\begin{minipage}{.21\columnwidth}\textbf{(d)}\end{minipage}
\begin{minipage}{.14\columnwidth}\textbf{(e)}\end{minipage}
\begin{minipage}{.19\columnwidth}\textbf{(f)}\end{minipage}
\begin{minipage}{.19\columnwidth}\textbf{(g)}\end{minipage}
\caption{
\textbf{(a)}
Part of a face-centered cubic (FCC) lattice (green and blue points).
\textbf{(b)}
Spherically-shaped i-surfaces (with radius 0.46) are placed at the vertices of the FCC lattice.
Here, we show seven i-surfaces positioned at the seven blue points in (a).
Qubits live on the lattice links (black) where two i-surfaces intersect.
(Some links are hidden behind the spheres.)
\textbf{(c-g)}
The resulting 3-cells (3D volume enclosed by i-surfaces).
The phase of the X-cube model on the resulting lattice is 3+1D $Z_2$ gauge theory.
Excitations of the 3-cell cubes (g) are $Z_2$ charges (not fractons).
The hopping operator is a product of $X$ operators on the two red links in (b),
  which excites the two nearby blue cubes ($Z_2$ charges).
}\label{fig:spheres}
\end{figure}

Now consider a lattice constructed from spherical i-surfaces.
We will place the spheres on a face-centered cubic (FCC) lattice (\sfigref{fig:spheres}{a-b}).
\foot{Placing the spheres on a cubic lattice results in four spheres intersecting at a single point,
  which violates the assumptions of our lattice construction.}
Remarkably, when the X-cube model is defined on this lattice
  \foot{In this case, the new loop terms in $H_\text{X-cube}$ (\eqnref{eq:Xcube H'}, \sfigref{fig:3cell}{b-c}) are actually optional.}
  the phase is equivalent to 3+1D $Z_2$ gauge theory (which can be described by BF theory or 3+1D toric code).
Indeed, we have checked that the ground state degeneracy is equal to 8 on this lattice lattice with periodic boundary conditions,
  as expected for 3+1D $Z_2$ gauge theory. \citefoot{foot:GSD}
(The unit cell is composed of 48 links, on which the Pauli operators reside.)

The important physics results from the fact that the lattice curvature allows
  the excitations of the 3-cell operators to be fully mobile (charge excitations in $Z_2$ gauge theory),
  rather than immobile fractons which is the case on a cubic lattice.
More generally, this tends to occur when many triples of i-surfaces intersect at multiple points.

An excitation of a single cross operator (\sfigref{fig:Xcube}{b}) still behaves similar to a dimension-1 particle;
  however it is localized to a finite-sized loop around a sphere,
  and is therefore effectively confined.
Certain (somewhat complicated) closed loops of cross-operator excitations corresponds to a $Z_2$ gauge theory flux.
Logical operators are given in \refcite{XcubeRG},
  which are useful for understanding the flux excitation.


\begin{figure}
\includegraphics[width=.9\columnwidth]{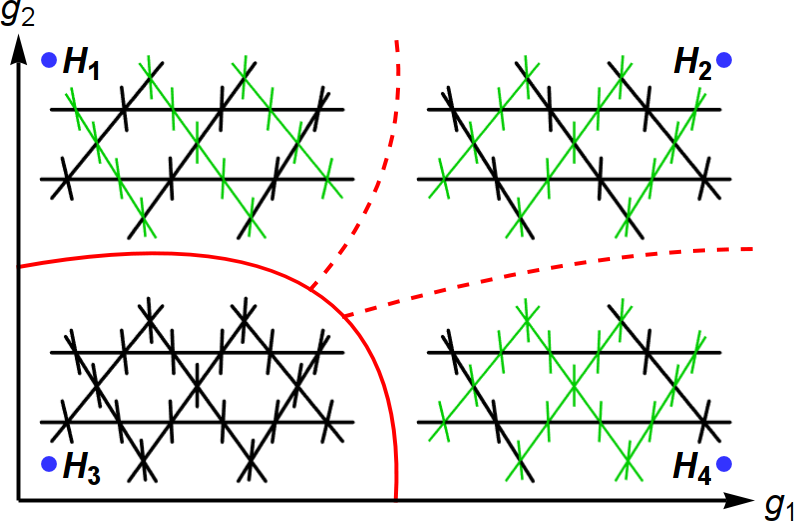}
\caption{
A schematic phase diagram of $H(g_1,g_2)$ (\eqnref{eq:Hg}),
  which interpolates between the four Hamiltonians $H_i$ ($i=1,2,3,4$) at the four corners (blue dots).
Each $H_i$ consists of the X-cube model defined on the black lattice (next to each $H_i$),
  while the qubits on the green links are trivially gapped out as in \eqnref{eq:Hg}.
The X-cube model is sensitive to the geometry of the lattice on which it is defined (black lattice),
  and as a result, each $H_i$ belongs to a different gLU-phase and the four $H_i$ are separated by phase transitions (solid and dotted red lines).
This physics is unique to fracton order;
  if we placed a liquid topological order (e.g. 3d toric code) on the black lattice (instead of the X-cube model),
  then the above phase transitions would not be necessary.
However, as explained in \secref{sec:isometry}, if lattice transformations are allowed during the Hamiltonian interpolation,
  then the phase transitions along the dotted red lines are not necessary,
  which leads to a new definition of phase of matter.
}\label{fig:phases}
\end{figure}

\mysec{Different Phases} \label{sec:phases}
We have shown that lattice geometry greatly affects the topological excitations, ground state degeneracy (GSD), and phase of the X-cube model.
In this section, we will discuss more subtle ways that the lattice geometry affects the phase of matter (\figref{fig:phases}),
  which will motivate a new definition of phase in \secref{sec:isometry}.

We will begin by studying the X-cube model using the definition of phase of matter defined in \refcite{XiePhases}.
In \refcite{XiePhases}, ground states of gapped local Hamiltonians are grouped into equivalence classes,
  which are interpreted as phases of matter.
The ground states of $H(0)$ and $H(1)$ are in the same phase if there exists an adiabatic evolution of Hamiltonians $H(g)$ such that $H(g)$ is gapped for all $0 \leq g \leq 1$
  (i.e. no phase transition occurs)
  where $g$ parametrizes the coupling constants in $H(g)$.
Equivalently, two states are in the same phase if they can be equated by a generalized local unitary transformation (gLU). \citefoot{foot:gLU}
We will refer to phases under this classification as gLU-phases
  in order distinguish them from the coarser classification of phase that we will propose in \secref{sec:isometry}.

We will argue that different lattice geometries result in different gLU-phases using two different kinds of arguments.
The first argument is to consider the GSD.
If two Hamiltonians have different GSD,
  then the Hamiltonians must be separated by a gap closing (i.e. a phase transition) so that the GSD can change
  (under an adiabatic evolution of the Hamiltonian $H(g)$).
This implies that e.g. the X-cube model on the following lattices must correspond to different gLU-phases:
  a cubic lattice, a stacked kagome lattice, or a cubic lattice with a larger unit cell (\figref{fig:phases}).

To make this explicit, we can consider an arbitrary interpolation $H(g_1,g_2)$ between four Hamiltonians $H_i$ with $i=1,2,3,4$:
\begin{align}
\begin{split}
  H_i &= H_\text{X-cube}(\cK) \label{eq:Hg}\\
      &\quad  - \sum_{\ell,\ell' \in \text{splitted }\cK} X_\ell X_{\ell'} - \sum_{\ell \in \cP} Z_\ell
\end{split} \\
  H(g_1,g_2) &\text{ interpolates between } H_i \text{ with } i=1,2,3,4 \nn
\end{align}
Each $H_i$ is defined on a stacked kagome lattice and consists of the X-cube model defined on the black lattice (next to each $H_i$ in \figref{fig:phases}),
  while the qubits on the green links are trivially gapped out.
The second term is not conceptually important and will be discussed in the next paragraph.
For a periodic stacked kagome lattice of length $L$,
  $H_1$ and $H_2$ describe the X-cube model on different slanted cubic lattices and with a $\text{GSD}=2^{6L-3}$.
$H_3$ describes the X-cube model on a stacked kagome lattice with $\text{GSD}=2^{8L-3}$.
And $H_4$ describes the X-cube model on a slanted cubic lattice with a larger unit cell and $\text{GSD}=2^{5L-3}$. \citefoot{foot:GSD}
Thus, $H_1$ (and $H_2$), $H_3$, and $H_4$ must belong to different gLU-phases since they have different GSD.
  \foot{Interestingly, in \refcite{XcubeRG} it is shown that a local unitary transformation can connect $H_2$ to the union of $H_4$ and decoupled stacks of toric code.}

\begin{figure}
\includegraphics[width=.5\columnwidth]{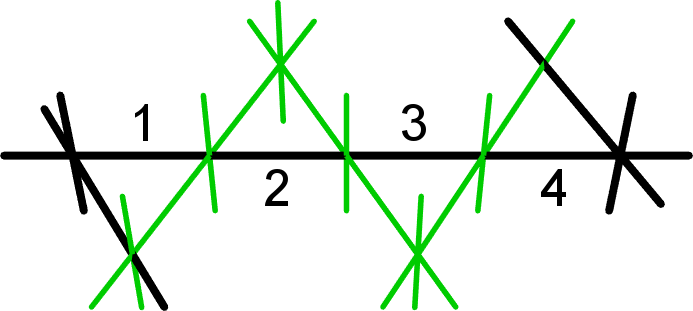}
\caption{
Vertices of the black slanted cubic lattices in \figref{fig:phases} can be separated by two or four black links.
In order to define an X-cube model on these black lattices with extra degrees of freedom,
  we need to add an additional term to $H_i$ (\eqnref{eq:Hg}),
  which takes the following form for the above for labeled links:
  $\sum_{\ell,\ell' \in \text{splitted }\cK} X_\ell X_{\ell'} = X_1X_2 + X_1X_3 + X_1X_4 + X_2X_3 + X_2X_4 + X_3 X_4$. 
}\label{fig:split}
\end{figure}

There is a minor subtlety regarding the definition of $H_\text{X-cube}(\cK)$ due to the fact that
  pairs of neighboring vertices in the black slanted cubic lattices (in \figref{fig:phases}) are split into multiple links (\figref{fig:split}) on the stacked kagome lattice.
In $H_\text{X-cube}(\cK)$, the 3-cell operators are products of $Z$ operators on all of the black links on the edge of a 3-cell.
The cross operators are still products of exactly four $X$ operators on four black links neighboring a vertex.
The second term in \eqnref{eq:Hg} sums over all pairs of links that are between two neighboring vertices in a black lattice;
  see \figref{fig:split} for an example.
We emphasize that $H_1$ is in the same phase as the X-cube model defined on a slanted cubic lattice like the one that $H_1$ is defined on,
  but without the green links or multiple black links between neighboring vertices;
  $H_1$ just includes extra qubit degrees of freedom which are either gapped out (as in the green links) or (trivially) sewed in using a local unitary (as in the splitted black links).

\begin{figure}
\includegraphics[width=.75\columnwidth]{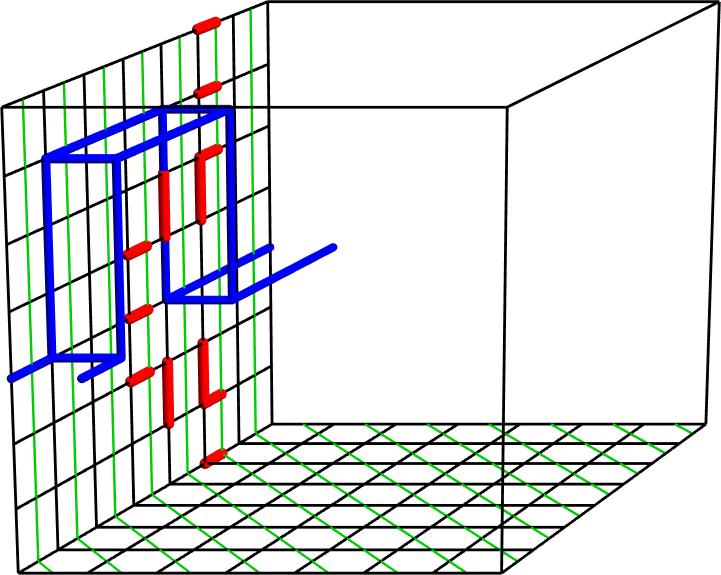}
\caption{
A 't Hooft loop (red) and paired Wilson loop (blue), which both commute with $H_1$ (\figref{fig:phases}) (but not $H_2$).
The 't Hooft and paired Wilson loop operators anticommute and are products of $X$ and $Z$ operators on the red and blue links, respectively.
The above loop operators each transport dimension-2 particles around a periodic direction via a non-linear path.
$H_2$ does not have topological excitations that can traverse the same plane
  (which becomes a precise statement in the limit of large system size).
In \appref{app:rotated ops}, we use this fact to argue that $H_1$ and $H_2$ correspond to different phases of matter.
}\label{fig:dim2ops}
\end{figure}

The two different slanted cubic lattices ($H_1$ and $H_2$) also belong to different gLU-phases.
Again, this can be shown using certain choices of periodic boundary conditions for which these Hamiltonians have different GSD \citeapp{app:rotated GSD}.
However, there is also a physical reason for a difference in gLU-phase.
The dimension-2 particles (i.e. pairs of fractons or dimension-1 particles) are bound to a i-surface
  (plane of the black cubic lattice in this case),
  and these i-surfaces are different in $H_1$ and $H_2$.
If we consider a 't Hooft loop or paired Wilson loop formed by these dimension-2 particles,
  then there are loop operators (\figref{fig:dim2ops}) that only exist in $H_1$ or $H_2$, but not both.
In \appref{app:rotated ops}, we use this intuition to show more formally that in general it is not possible to
  relate the ground states of $H_1$ and $H_2$ by a local unitary transformation.
$H_1$ and $H_2$ can be related by a lattice rotation, but this is not a local unitary transformation.
Thus, the X-cube model on a rotated lattice can result in a different gLU-phase of matter from the X-cube model on the un-rotated lattice.
\foot{A translated a lattice, by e.g. half of a unit cell, does not result in a different phase
  since a lattice translation can be expressed as a generalized local unitary (gLU) transformation.}

\subsection{Coarser Fracton Phases}
\label{sec:isometry}

In \refcite{XiePhases}, phases of gapped local Hamiltonians were defined as being separated by phase transitions;
  equivalently, different gLU-phases can not be connected by a generalized local unitary (gLU) transformation \citefoot{foot:gLU}.
However, we argued that the X-cube model on the different black lattices in \figref{fig:phases} correspond to different gLU-phases of matter that must be separated by phase transitions.
This may be unsettling since the lattices used to define $H_1$, $H_2$, and $H_4$ in \figref{fig:phases} only differ by rotation or rescaling of the unit-cell size.
This motivates a new definition of phase of matter where
  the gLU is composed (via function composition) with a \emph{quasi-isometry}.

A quasi-isometry is a spatial transformation that preserves long-distance structure, such as locality.
A quasi-isometry is coarsely (i.e. up to a bounded error) one-to-one and onto.
Importantly, a quasi-isometry preserves locality (of local Hamiltonians and local unitary transformations)
  since a quasi-isometry $f$ must satisfy $|f(x) - f(y)| \leq K |x-y| + A$ (for constant $K,A>0$),
  which implies that nearby points can not be mapped to far away points.
A quasi-isometry does not need to preserve short-distance details;
  e.g. a quasi-isometry can be discontinuous and is perfectly well-defined on a lattice.
These properties are desirable since we are interested in long-distance physics.
A homeomorphism (between path-connected metric spaces) with bounded derivatives is a quasi-isometry
  that also preserves short-distance structure.
See \appref{app:locality} for more details.
Rotations, translations, and scale transformations are the most important examples of quasi-isometries in this work.

We propose a new definition of phase where two states are in the same phases if and only if they can be
  equated by the composition of a gLU and a quasi-isometry.
This definition of phase intuitively groups $H_1$, $H_2$, and $H_4$ (\figref{fig:phases}) into the same phase,
  but places $H_3$ into a separate phase.

Recall that phases of gapped Hamiltonians are separated by phase transitions under adiabatically varying the Hamiltonian.
If the positions of the degrees of freedom (qubits in this work) are stationary under the adiabatic evolution,
  then gLU alone defines the corresponding notion of phase. 
However, if the positions of the degrees of freedom can be changed during the adiabatic evolution
  (e.g. by rotating or applying strain or pressure to a crystal in a lab),
  then the composition of a gLU with a quasi-isometry defines the correct notion of phase.
That is because in this case,
  the ground states are separated by phase transitions if and only if they can not be related by the composition of a gLU with a quasi-isometry.

\mysec{Conclusion}
We have shown that nearly all important characterizations of the long-distance physics of the X-cube model depend on lattice geometry.
These characterizations include:
  the mobility restrictions of the topological excitations and braiding operators,
  ground state degeneracy,
  and the phase of matter.
We emphasize that the long-distance physics of liquid topological order is blind to the short-distance lattice geometry
  since none of the above characteristics of liquid topological order depend on the geometry of the lattice.
The lattice geometry dependence of fracton order is ultimately the reason why our new definition of phase of matter (\secref{sec:isometry}) was necessary.
Thus, we propose that fracton orders should be called fracton \emph{geometric order}
  to emphasize the important role played by geometry.
(Referring to fracton orders as ``fracton topological order'' is misleading since these phases are not topologically invariant.)

In \secref{sec:isometry}, we made use of quasi-isometries to define phases of matter.
The paradigm that we are applying here is that metric spaces and quasi-isometries are useful mathematical tools when one is interested in long-distance physics
    but starting with a short-distance model.
Quasi-isometries become unnecessary after coarse-graining (under renomalization group) a model to its low-energy and long-distance effective field theory,
  for which the short-distance details have been thrown away and no longer need to be explicitly ignored by using quasi-isometries.
For example, after coarse-graining toric code to BF theory
  \foot{The relation between toric code \cite{KitaevToric} and BF theory \cite{BFTheory} is reviewed in the appendix of \refcite{fractonQFT}.},
  one finds that metric spaces, distance metrics, and quasi-isometries can be replaced by coarser concepts:
  topology, continuity, and continuous functions.
A similar paradigm using quasi-isometries was recently applied in \refcite{PhysicsWithoutPhysics,GeoRG} in order to connect lattice physics to quantum field theories within an ``it from qubit'' perspective.

Mathematically, quasi-isometries are particularly important in the study of geometric group theory.
Geometric group theory has been observed \cite{Bacon_2003} to have connections to the Solovay-Kitaev theorem \cite{Kitaev_1997,Solovay2000},
  which is an important result in the theory of quantum computation.
Roughly, the theorem says that if a set of operators generates a dense subset of $SU(2)$,
  then any operator in $SU(2)$ can be efficiently obtained within an accuracy $\epsilon$ by
  taking a product of only $\sim \log^4(1/\epsilon)$ operators in the generating set \cite{Dawson_Nielsen_2005}.
It may be interesting to use geometric group theory to understand the geometry of the ground state degeneracy of fracton orders.

We have only briefly studied examples of how geometry affects the physics of the X-cube model.
A more complete and general understanding of the generic mathematical structure would be very desirable.
For example, it would be interesting to study the infinite hierarchy of X-cube models that results from our lattice construction via increasing the number of stacks of i-surfaces.
  (We only considered three or four stacks, which resulted in cubic and stacked kagome lattices, respectively.)
Furthermore, if the X-cube model is the simplest example of fracton order in the same way that toric code is the simplest example of topological order,
  more interesting geometric physics may emerge in other fracton models,
  similar to how more interesting topological invariants result from more exotic models of (liquid) topological order \cite{PutrovBraiding}.

Previously-proposed emergent gravity models \cite{Gu_Wen_2006,Gu2012,Xu_2006,Xu2006,Xu2010} were later discovered to actually be gapless fracton models \cite{PretkoGravity}.
And recently, a gravity-like attraction mechanism between fractons was discovered \cite{PretkoGravity} in gapless $U(1)$ fracton models \cite{Rasmussen2016,PretkoU1}
  (although the attractive force is only long ranged if the model has gapless fracton dipoles).
These gapless $U(1)$ fracton models therefore appear to be simplified versions of a gravity-like model.
The gapped $Z_N$ fracton models discussed in this work are the discrete analogs of the $U(1)$ fracton models.
It would therefore be interesting to study how the geometry-dependent physics discussed in this work applies to the $U(1)$ fracton models,
  and how the gravity-like connections of the $U(1)$ fracton models may apply to the gapped $Z_N$ fracton models.

\acknowledgments
We thank Wilbur Shirley, Xie Chen, Zhenghan Wang, Michael Pretko, Tim Hsieh, Wonjune Choi, and Dave Aasen for many helpful and encouraging discussions.
This work was supported by the NSERC of Canada and the Center for Quantum Materials at the University of Toronto.

\bibliography{XcubeLattice14}

\begin{thebibliography}{65}
\bibitem [{foo({\natexlab{a}})}]{foot:daggers}%
  \BibitemOpen
  \href@noop {} {} ({\natexlab{a}}),\ \bibinfo {note} {{I}n \cite{VijayXCube},
  a $Z_N$ generalization of the X-cube model was given. However, their $Z_N$
  model only generalizes to lattices with bipartite 3-cells. That is, the
  vertices of every 3-cell must form a bipartite lattice. In our figures, we
  present an alternative $Z_N$ generalization which does not require bipartite
  3-cells. However, our generalization has a peculiarity where odd integer
  system sizes can result in a reduced GSD. For simplicity, we will focus on
  the $Z_2$ model in the main text, for which $X^\dagger = X$.}\BibitemShut
  {Stop}%
\bibitem [{foo({\natexlab{b}})}]{foot:GSD}%
  \BibitemOpen
  \href@noop {} {} ({\natexlab{b}}),\ \bibinfo {note} {{G}round state
  degeneracies were calculated using a method equivalent to the method
  described in Appendix B of \refcite{MaLayers}.}\BibitemShut {Stop}%
\bibitem [{foo({\natexlab{c}})}]{foot:gLU}%
  \BibitemOpen
  \href@noop {} {} ({\natexlab{c}}),\ \bibinfo {note} {{L}ocal unitary
  transformations can be generated by a finite time evolution of a local
  time-dependent Hamiltonian. When considering an adiabatic evolution of a
  Hamiltonian $H(g)$ from $H(0)$ to $H(1)$, it is allowed (and sometimes
  necessary) to add extra degrees of freedom to $H(g)$ (and thus also to $H(0)$
  and $H(1)$) as long as these extra degrees of freedom are gapped and
  decoupled when $g=0$ or $g=1$. Analogously, a generalized local unitary (gLU)
  \cite{XiePhases} is a local unitary transformation that can also add or
  remove degrees of freedom that are in a direct product state.}\BibitemShut
  {Stop}%
\end{thebibliography}

\appendix

\section{Rotated Lattice Arguments}
\label{app:rotated}

In this appendix, we will argue more thoroughly that the X-cube model on lattices that only differ by a rotation can result in different gLU-phases of matter.
As a concrete example, we will show that the ground states of $H_1$ and $H_2$ (\figref{fig:phases}) belong to different gLU-phases.
That is, we will argue that $H_1$ and $H_2$ must be separated by a phase transition (assuming immobile qubits)
  and their ground states can not be related by a generalized local unitary (gLU) transformation.

\subsection{Degeneracy Argument}
\label{app:rotated GSD}

\begin{figure}
\includegraphics[width=.6\columnwidth]{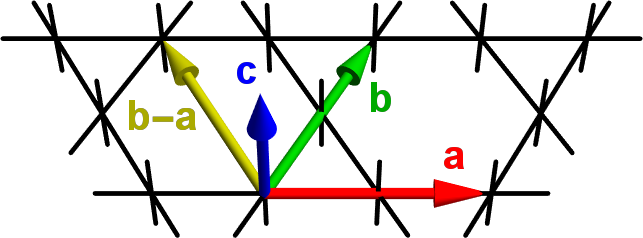}
\caption{
Lattice vectors used to define the periodic boundary conditions in \eqnref{eq:periodic} and (\ref{eq:periodic'}).
}\label{fig:latticeVec}
\end{figure}

First, we will give an example of a certain periodic boundary condition for which the two Hamiltonians ($H_1$ and $H_2$) have a different ground state degeneracies (GSD).
A periodic lattice can be defined by imposing a periodic equivalence of lattice points.
A typical choice for a lattice of lengths $(L_1,L_2,L_3)$ is to equate each point $\mathbf{x}$ as follows:
\begin{equation}
  \mathbf{x} \equiv \mathbf{x} + L_1 \mathbf{a} \equiv \mathbf{x} + L_2 \mathbf{b} \equiv \mathbf{x} + L_3 \mathbf{c} \label{eq:periodic}
\end{equation}
where $\mathbf{a}$, $\mathbf{b}$, and $\mathbf{c}$ are lattice vectors.
With the lattice vectors shown in \figref{fig:latticeVec},
  the above periodic boundary conditions result in a $\log_2 \text{GSD}=2L_1+2L_2+2L_3-3$ for both $H_1$ and $H_2$.

However, if we instead choose
\begin{align}
  \mathbf{x} &\equiv \mathbf{x} + \mathbf{b} + L_1 \mathbf{a} \equiv \mathbf{x} + L_2 \mathbf{b} \equiv \mathbf{x} + L_3 \mathbf{c} \label{eq:periodic'}
\end{align}
then $H_1$ has $\log_2 \text{GSD} = 2L_2 + 2L_3$ (for even $L_1$) while $H_2$ instead has a much smaller $\log_2 \text{GSD} = 2 + 2 L_3$ (for even $L_1$ and $L_2$) \citefoot{foot:GSD}.
The reduced GSD occurs because the Wilson loops (\sfigref{fig:loopsAndDegen}{a}) along certain directions get merged into a single Wilson loop due to the shifted periodic boundary condition.
Since $H_1$ and $H_2$ have different GSD on the same lattice,
  they must be separated by a phase transition
  (if we assume that the qubits are immobile under adiabatic Hamiltonian evolution).

\subsection{Logical Operator Argument}
\label{app:rotated ops}

As an alternative argument, we will derive a contradiction by assuming that the ground states of $H_1$ and $H_2$ (\eqnref{eq:Hg}) can be related by a local unitary transformation
  (in the sense of \refcite{XiePhases}).
We will work in the limit of an infinitely large lattice so that the concept of local operators is well-defined.
This will be an argument (not a proof) since we will apply the physics knowledge that $H_2$ does not have mutual semion topological excitations that can traverse the loops in \figref{fig:dim2ops}.
However, we expect that this could also be proven.

To derive a contradiction, consider a ground state $\ket{\psi_1}$ of $H_1$ that is an eigenstate of the 't Hooft loop $T_1$ in \figref{fig:dim2ops},
  which wraps a dimension-2 particle (composed of a pair of fractons) around a periodic direction.
We will also consider the paired Wilson loop operator $W_1$ in \figref{fig:dim2ops},
   which wraps a dimension-2 particle (composed of a pair of dimension-1 particles) around the same plane.
The algebra of these operators is
\begin{align}
  T_1 \ket{\psi_1} &= \ket{\psi_1} &
  T_1 W_1 = - W_1 T_1
\end{align}

To derive a contradiction, suppose that $H_2$ has a ground state $\ket{\psi_2}$
  that is related to $\ket{\psi_1}$ by a local unitary transformation $U$ via $\ket{\psi_2} = U \ket{\psi_1}$.
We can then define the transformed operators
\begin{align}
  T_2 &\equiv U T_1 U^{-1} & W_2 &\equiv U W_1 U^{-1}
\end{align}

$\ket{W_2 \psi_2} \equiv W_2 \ket{\psi_2}$ must also be a ground state of $H_2$.
This is because
\begin{align}
\begin{split}
  \brakets{H_2^n}{    \psi_2} &= \underset{\quad\scalebox{1.2}{\rotatebox{90}{$=\;$}}}
                                  {\brakets{U^{-1} H_2^n U}{    \psi_1}} \label{eq:GS}\\
  \brakets{H_2^n}{W_2 \psi_2} &=  \brakets{U^{-1} H_2^n U}{W_1 \psi_1}
\end{split}
\end{align}
where we have raised $H_2$ to a positive integer power $n$.
The right hand sides of the above two lines must be equal since
  $\ket{\psi_1}$ and $\ket{W_1 \psi_1} \equiv W_1 \ket{\psi_1}$ are ground states of $H_1$ (which is topologically ordered)
  and must therefore be indistinguishable by local operators
\foot{Recall that if it were possible to distinguish $\ket{\psi_1}$ and $\ket{W_1 \psi_1}$ using the expectation value of a local operator $H'$,
  then $H'$ could be added to the Hamiltonian $H_1$ with an arbitrarily small coefficient and split the degeneracy between $\ket{\psi_1}$ and $\ket{W_1 \psi_1}$,
  which is impossible because the ground state degeneracy of the X-cube model on a periodic cubic lattice is robust to perturbations \cite{VijayXCube}.},
  and $U^{-1} H_2^n U$ is a local operator (since $H_2$ and $U$ are local).
Using $n=1$ and $n=2$, \eqnref{eq:GS} shows that $\ket{\psi_2}$ and $\ket{W_2 \psi_2}$ have the same energy expectation value and energy uncertainty of $H_2$.
Since $\ket{\psi_2}$ is a ground state of $H_2$ with zero energy uncertainty,
  $\ket{W_2 \psi_2}$ is therefore also a ground state of $H_2$.

Therefore, $T_2$ and $W_2$ are logical operators that act on the ground states $\ket{\psi_2}$ and $\ket{W_2 \psi_2}$ of $H_2$ with the following algebra
\begin{align}
  T_2 \ket{\psi_2} &= \ket{\psi_2} & T_2 W_2 = - W_2 T_2 \label{eq:TWr}
\end{align}
since $\ket{\psi_2} = U \ket{\psi_1}$.
But since $U$ is local, $T_2$ and $W_2$ must only act on the qubits near the (red and blue) loops drawn in \figref{fig:dim2ops}.
However, $H_2$ does not have logical operators obeying \eqnref{eq:TWr} that only act in this region.
This is because the presence of these string logical operators would imply the existence of topological excitations
  that can move along the loops drawn in \figref{fig:dim2ops}.
However, no such excitations exist for $H_2$.
We have thus derived a contradiction by assuming that the ground states of $H_1$ and $H_2$ can be related by a local unitary transformation.
(Although we did not consider \emph{generalized} local unitary (gLU) transformations,
  which can add and remove qubits,
  we do not expect this to affect the result of the argument.)
Therefore, the ground states of $H_1$ and $H_2$ can not be related by a local unitary transformation
  and must therefore correspond to different gLU-phases of matter.

Note that in order to derive a contradiction, it was essential that we assumed that $U$ is a \emph{local} transformation.
In particular, $U$ can not be a $2\pi/3$ lattice rotation operator and thus $T_2$ and $W_2$ can not simply be equal to $T_1$ and $W_1$ rotated by $2\pi/3$.
Since lattice rotations are an example of a quasi-isometry,
  this argument does not apply to our new definition of phase in \secref{sec:isometry}.

\section{Quasi-isometry and Locality}
\label{app:locality}

In this appendix, we will review the mathematical notion of quasi-isometry and prove that quasi-isometries preserve the locality of local Hamiltonians and local unitary transformations.

\begin{figure}
\includegraphics[width=.7\columnwidth]{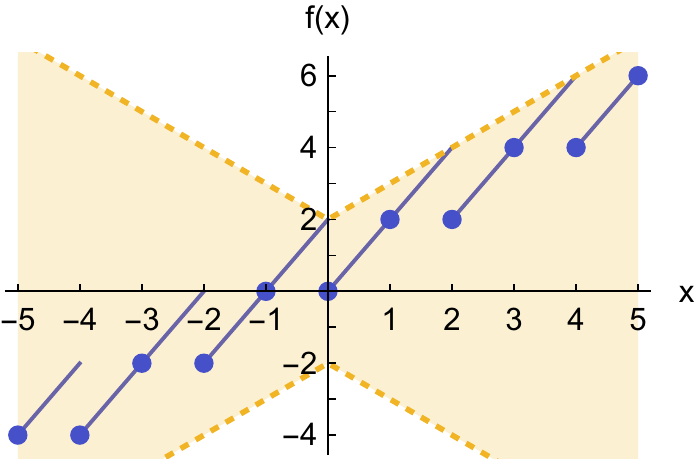}
\caption{
\textbf{(blue)} A quasi-isometry $f(x) = 2x - 2\;\text{floor}(x/2)$ which can be viewed on the continuum ($f : \RR \to \RR$) or a lattice ($f : \ZZ \to \ZZ$).
Note that $f$ is neither continuous nor one-to-one (nor onto when $f : \ZZ \to \ZZ$);
  these properties are not required of quasi-isometries.
$f(x)$ happens to be its own quasi-inverse (with $A=4$ in \eqnref{eq:inverse}).
\textbf{(yellow)} $f(x)$ satisfies the coarse-Lipschitz property (\eqnref{eq:KA}) with $K=1$ and $A=2$.
The shaded yellow region corresponds to the bound (resembling a light cone) that \eqnref{eq:KA} implies on $f(x)$ when $y=0$.
}\label{fig:isometry}
\end{figure}

A quasi-isometry is a function between metric spaces that is coarse Lipschitz and has a coarse-Lipschitz quasi-inverse.
To be precise, let $X$ and $X'$ be metric spaces with distance metrics $d_X : X \times X \to \RR$ and $d_{X'} : X' \times X' \to \RR$.
We will abuse notation and simply write $|x-y|$ instead of $d_X(x,y)$ or $d_{X'}(x,y)$
  (even though subtraction is not defined on a generic metric space).
A function $f : X \to X'$ is coarse-Lipschitz if
\begin{equation}
  |f(x) - f(y)| \leq K |x-y| + A \label{eq:KA}
\end{equation}
for all $x,y \in X$.
(If $A=0$, then $f$ is Lipschitz continuous.)
A function $g : X' \to X$ is a quasi-inverse of $f$ if there exists $A>0$ such that
\begin{equation}
  |g(f(x)) - x| \leq A \quad\text{and}\quad |f(g(x')) - x'| \leq A \label{eq:inverse}
\end{equation}
for all $x \in X$ and $x' \in X'$.
$g$ is a coarse inverse of $f$ in the sense that
  $g$ inverts $f$ up to a bounded error, $A$.
See \figref{fig:isometry} for an example.

As an aside, quasi-isometries can be related to other kinds of functions in certain limits:
A quasi-isometry with $A=0$ (in \eqnref{eq:KA} and \eqref{eq:inverse}) is also a uniform homeomorphism,
  which is a homeomorphism that is uniformly continuous and has a uniformly continuous inverse.
All uniform homeomorphisms between path-connected metric spaces are quasi-isometries.
A quasi-isometry with $K=1$ and $A=0$ is an isometry,
  which is a distance-preserving transformation between metric spaces.

We define a local Hamiltonian to be a sum of operators where the norm of each operator decreases exponentially with the diameter of its support.
More precisely, $\hat{H}$ is a local Hamiltonian if it can be expressed as $\hat{H} = \sum_\alpha \hat{O}_\alpha$
  with constants $E>0$ and $\xi>0$ such that
\begin{equation}
  ||\hat{O}_\alpha|| \leq E \, e^{-\text{diam}(\hat{O}_\alpha)/\xi} \label{eq:local}
\end{equation}
  where $||\hat{O}_\alpha||$ denotes the operator norm of $\hat{O}_\alpha$ and
\begin{equation}
  \text{diam}(\hat{O}_\alpha) = \text{diam}(X_\alpha) = \max_{x,y \in X_\alpha} |x-y| \label{eq:diam}
\end{equation}
is the diameter of the support of the operator $\hat{O}_\alpha$,
  where the support of $\hat{O}_\alpha$ is the set of lattice sites $X_\alpha$ that $\hat{O}_\alpha$ acts on.
(We will place hats above operators in this appendix.)
As an example, $\hat{H} = \sum_{ij} e^{-|i-j|} \hat{Z}_i \hat{Z}_j$ is a local Hamiltonian with $E=\xi=1$.

We consider quasi-isometries since unlike generic spatial transformations,
  quasi-isometries will preserve the locality of local Hamiltonians and local unitary transformations.
Furthermore, quasi-isometries have the desired property that they are not sensitive to short-distance details.
Quasi-isometries are introduced in this work so that we can be very precise about what kind of ``nice'' spatial transformations are allowed in our new definition of phase of matter,
  which applies to phases with or without fractons.

When we consider the composition of a quasi-isometry with a generalized local unitary (gLU) transformation,
  we will want to preserve the Hilbert space and locations of the degrees of freedom.
For example, the quasi-isometry $f(x) = 2x$
  will rescale a lattice of qubits at $x \in \mathbb{Z}$ to a lattice with qubits at even integers ($x \in 2\mathbb{Z}$).
(A similar rescaling relates the black lattices for $H_2$ and $H_4$ in \figref{fig:phases}.)
In order to preserve the Hilbert space,
  we must make use of the \emph{generalzied} local unitary (gLU) transformation \citefoot{foot:gLU}
  by adding back qubits at the odd integer positions ($x \in 2\mathbb{Z}+1$).
Since the Hilbert space and position of degrees of freedom are preserved,
  the physical notion of distance (i.e. the distance metric) can also be preserved.

As a non-example, although it is a homeomorphism, $f(x) = x^3$ is not a quasi-isometry.
This is desirable because on a lattice (where $f : \ZZ \to \ZZ$) and for large $x \gg 1$,
  $f$ will map neighboring qubits (at $x$ and $x+1$) to distant locations ($x^3$ and $(x+1)^3$)
  which will be separated by many qubits after adding back the missing qubits (at $x^3+1,x^3+2,\cdots,(x+1)^3-1$) with a gLU.

To show that quasi-isometries preserve the locality of local Hamiltonians,
  consider a quasi-isometry $f$ and a local Hamiltonian $\hat{H} = \sum_\alpha \hat{O}_\alpha$.
Let $K,A>0$ obey \eqnref{eq:KA}
  and choose $E,\xi>0$ that satisfy \eqnref{eq:local}.
After applying the quasi-isometry $f$,
  each operator $f(\hat{O}_\alpha)$ will have a diameter with the following upper bound:
\begin{align}
  \text{diam}(f(\hat{O}_\alpha))
    &=    \max_{x,y \in X_\alpha} |f(x)-f(y)| \label{eq:diam1} \\
    &\leq \max_{x,y \in X_\alpha} K |x-y| + A \label{eq:bound1} \\
    &=    K \text{diam}(\hat{O}_\alpha) + A \label{eq:diam2}
\end{align}
\eqnref{eq:diam1} and \eqref{eq:diam2} follow from the definition of the diameter of an operator (\eqnref{eq:diam}),
  while \eqnref{eq:bound1} follows from the coarse-Lipschitz property (\eqnref{eq:KA}).
Thus, after the quasi-isometry $f$ is applied to the Hamiltonian $\hat{H}$,
  the resulting Hamiltonian $f(\hat{H})$ is local since
\begin{align}
  ||f(\hat{O}_\alpha)||
    &=    ||\hat{O}_\alpha|| \label{eq:fO}\\
    &\leq E  \, e^{-\text{diam}(  \hat{O}_\alpha) /\xi } \label{eq:local1}\\
    &\leq E' \, e^{-\text{diam}(f(\hat{O}_\alpha))/\xi'} \label{eq:diam3}
\end{align}
where $E' = E \, e^{A/\xi'}$ and $\xi' = K \xi$.
\eqnref{eq:fO} follows since the quasi-isometry $f$ does not change the norm of an operator.
\eqnref{eq:local1} follows from the definition of a local Hamiltonian (\eqnref{eq:local}).
\eqnref{eq:diam3} follows from \eqnref{eq:diam2}.

Since local unitary transformations are defined in terms of time-dependent local Hamiltonians \citefoot{foot:gLU}
  and quasi-isometries preserve the locality of local Hamiltonians,
  this implies that quasi-isometries also preserve the locality of local unitary transformations.

\end{document}